\pdfoutput=1
\documentclass[aps,pra,reprint,superscriptaddress]{revtex4-2}
\usepackage[english]{babel}

\makeatletter
\def\bbl@set@language#1{%
    \edef\languagename{%
        \ifnum\escapechar=\expandafter`\string#1\@empty
        \else\string#1\@empty\fi}%
    \@ifundefined{babel@language@alias@\languagename}{}{%
        \edef\languagename{\@nameuse{babel@language@alias@\languagename}}%
    }%
    \select@language{\languagename}%
    \expandafter\ifx\csname date\languagename\endcsname\relax\else
    \if@filesw
    \protected@write\@auxout{}{\string\select@language{\languagename}}%
    \bbl@for\bbl@tempa\BabelContentsFiles{%
        \addtocontents{\bbl@tempa}{\xstring\select@language{\languagename}}}%
    \bbl@usehooks{write}{}%
    \fi
    \fi}
\newcommand{\DeclareBabelLanguageAlias}[2]{%
    \global\@namedef{babel@language@alias@#1}{#2}%
}
\makeatother
\DeclareBabelLanguageAlias{en}{english}

\usepackage{hyperref}
\hypersetup{
    colorlinks=true,
    allcolors=red,
    citecolor=blue,
    urlcolor=blue
}
\usepackage{amssymb,mathtools}
\usepackage{physics}
\usepackage{standalone}
\usepackage{easyReview}

\newcommand{\Figref}[2][]{\hyperref[#2]{\ref*{#2}\ifthenelse{\equal{#1}{}}{}{\thinspace#1}}}

\begin{document}

\title{Bisingular surface polaritons at the interface of two uniaxial media}

\author{K. Yu. Golenitskii}
\email[]{golenitski.k@mail.ioffe.ru}
\affiliation{Ioffe Institute, 194021 St.~Petersburg, Russia}
\affiliation{Skolkovo Institute of Science and Technology, 121205 Moscow, Russia}
\author{N. S. Averkiev}
\affiliation{Ioffe Institute, 194021 St.~Petersburg, Russia}

\date{\today}

\begin{abstract}
In some anisotropic bulk media (for example, biaxial weakly absorbing crystals) there are special directions along which the plane wave field distribution has a singular profile of the form \( \propto (\vb{n} \vb{r}) \exp(i q \vb{n} \vb{r}) \).
They are also known as Voigt waves.
Similar singular profiles also arise in the theory of surface electromagnetic waves in anisotropic media.
In this work we have considered surface polaritons at the interface of two, generally different, uniaxial media.
Optic axis of both media is parallel to the interface.
One of the specific solutions, called bisingular, greatly simplifies the dispersion equation for surface polaritons.
In this case the analytical solution in closed form is found and existence conditions have been determined.
It is shown that bisingular surface polariton exists only for certain angles between the optic axes, which are found from two cubic equations.
All parameters of the bisingular surface polariton depend only on permittivities and this angle.
If one medium is weakly anisotropic, or both media are almost the same then two angles exist.
In the general case of two arbitrary media there can be from two to six such angles.
\end{abstract}

\maketitle


\section{Introduction}
Anisotropic materials, in contrast to isotropic materials, offer significantly greater opportunities for controlling the polarization of light and distribution of electromagnetic fields.
The anisotropy of the medium leads to new effects that are impossible to achieve in isotropic media.
It is known that special directions can exist in an anisotropic medium called singular axes \cite{Pancharatnam1958LightPropagation} (or circular axes \cite{Khapalyuk1962TheoryCircular,Fedorov1963PropagationLight,Fedorov1976TeoriyaGirotropiia}).
The solution of Maxwell's equations in the form of a plane wave propagating in these directions is expressed as
\begin{equation}\label{eq:voigt-solution}
    \vb{E} = \qty[\vb{f}_0 + \vb{f}_1 \qty(\vb{n} \vb{r}) ]\, \mathrm{e}^{i q \vb{n} \vb{r} - i \omega t}.
\end{equation}
This was first noticed by Voigt \cite{Voigt1902BehaviourPleochroitic} for absorbing pleochroitic biaxial crystals.
A rigorous theoretical explanation was given by Pancharatnam \cite{Pancharatnam1958LightPropagation} and independently by Khapalyuk \cite{Khapalyuk1962TheoryCircular} (see also works \cite{Fedorov1963PropagationLight,Gerardin2001ConditionsVoigt} and review \cite{Mackay2023AnatomyVoigt}).
Singular waves have also been studied experimentally \cite{Brenier2015VoigtWave}.
Constant vectors \( \vb{f}_0 \) and \( \vb{f}_1 \) determine the polarization of the wave, \( \vb{n} \) is the unit vector of the propagation direction, \( q \) is the wavevector, \( \omega \) is the wave frequency.
It may seem that the second term in \eqref{eq:voigt-solution} corresponds to a nonphysical solution because it grows linearly with the coordinate.
This can be explained clearly by the fact that the wavevector \( q \) in absorbing crystals is a complex number, and \( \Im q \) always corresponds to attenuation along the direction of propagation.
Consequently, there is always an exponential decay that overwhelms the linear growth.
Singular axes do not exist in non-absorbing biaxial crystals \cite{Fedorov1976TeoriyaGirotropiia} because, in this case, we can always choose two orthogonal polarizations instead of one.
Generally, the existence of singular axes in a medium is related to the possibility of wedge refraction \cite{Alshits2004WedgeRefraction}, a phenomenon very close in meaning to conical refraction \cite{Turpin2016ConicalRefraction}.

In the theory of surface electromagnetic waves in an anisotropic medium, singular solutions similar to Eq.~\eqref{eq:voigt-solution} also arise.
However, in this case, Voigt-type solutions may exist in both the biaxial and uniaxial media \cite{Fedorov1976TeoriyaGirotropiia,Marchevskii1984SingularElectromagnetic,Mackay2019DyakonovVoigt,Zhou2019SurfaceplasmonpolaritonWave,Lakhtakia2020ElectromagneticSurface,Golenitskii2024AnisotropicSurface}.
Moreover, the media may be non-absorbing.
These singular surface waves are of interest for several reasons.
Firstly, the polarization of the wave is dependent on the coordinates, which is unusual in a non-gyrotropic medium.
This is less pronounced in surface waves compared to bulk waves.
Nevertheless, this suggests an intriguing situation that is worth examining more closely.
Secondly, the existence of a singular surface wave implies the existence of a simple (nonsingular) surface wave too.
The problem of surface wave propagation in two arbitrary anisotropic media is complex and often involves numerical modeling.
At best, the existence conditions and/or the angular domain of propagation are explicitly known.
The remaining parameters of the surface wave have to be found numerically.
In some special cases, it is possible to obtain a good approximate or even exact analytical solution owing to additional internal symmetries.
It is the singular form that make it possible to obtain the analytical results.

This study examines surface polaritons propagating along the interface of two arbitrary nonabsorbing uniaxial media.
An analytical solution is found when the polariton is singular in both media, which we call bisingular.
Some special cases were analyzed in detail.

\section{Model}
\begin{figure}
    \includegraphics[width=0.9\linewidth]{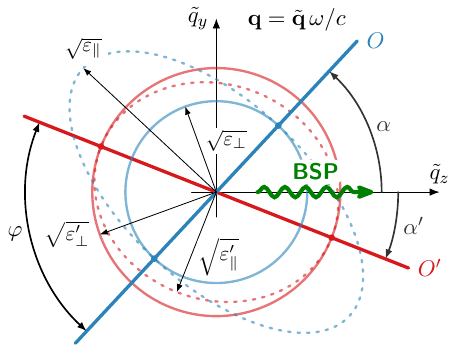}
    \caption{
        \label{fig:sketch}An example of mutual orientation of the optic axes of different uniaxial dielectric crystals.
        One of the crystals is positive uniaxial \(\varepsilon_\perp < \varepsilon_\parallel\), and the other is negative uniaxial \(\varepsilon'_\perp > \varepsilon'_\parallel \).
        \( (yz) \) is the interface plane between crystals.
        Bisingular surface polariton (BSP) propagates along \( z \) axis.
        Optic axes \( O \) and \( O' \) are parallel to the interface.
        Solid circles represent the boundaries of possible wave vectors of bulk ordinary waves.
        Similarly, dashed ellipses represent the boundaries of possible wave vectors of bulk extraordinary waves.
    }
\end{figure}

Let the interface between different uniaxial crystals be the plane \( x = 0 \) (Fig.~\ref{fig:sketch}).
Let us choose the directions of the other axes such that the surface polariton propagates along the \( z \) axis.
We use the prime sign, e.g. \( \alpha' \), to denote quantities related to the upper crystal \( x > 0 \) and no additional signs for quantities related to the bottom crystal \( x < 0 \).
The optic axes \( O \) and \( O' \) of both crystals are parallel to the interface, and the angle between them is \( \varphi \in (0, \pi)\).
In the chosen coordinates the electric and magnetic fields do not depend on \( y \); instead, the dielectric tensor \( \hat{\varepsilon} \) has off-diagonal elements \( \varepsilon_{yz} = \varepsilon_{zy} \).
Dielectric tensor in principal axes is diagonal \( \hat{\varepsilon} = \operatorname{diag} (\varepsilon_\perp, \varepsilon_\perp, \varepsilon_\parallel) \).
Let \( \alpha \) be the angle by which the upper crystal should be rotated such that the optic axis coincides with the \( z \) axis.
Angle \( \alpha' \) is defined similarly.
It is clear that it is sufficient to consider \( \alpha, \alpha' \in (-\pi/2, \pi/2) \).
Therefore, the angle between the optic axes is \( \varphi = \alpha - \alpha' \).

We seek Maxwell's equations in the form of a monochromatic plane wave propagating along the interface \( {\mathbf{E}, \mathbf{H} \propto \exp(i q z - i \omega t)} \).
After the substitution, the equations depend only on the \( x \) coordinate.
Only solutions localized near the interface, and decreasing with distance from it are of interest.
The non-singular general solution in each crystal is a linear combination of ordinary and extraordinary wave \cite{Averkiev1990ElectromagneticWaves}.
It is important that their localization constants are not equal in this case.
However, if
\begin{equation*}
    q^2 = \frac{\omega^2}{c^2} \frac{\varepsilon_\perp}{\cos^{2} \alpha}
\end{equation*}
and \( \alpha \neq 0 \) or \( \pi/2 \), then the general solution has a singular form.
Throughout the following it is assumed that both \( \alpha \) and \( \alpha' \) differ from \( 0 \) or \( \pi / 2 \).

The electric field for a singular surface polariton in the bottom crystal \( x < 0 \) can be written as
\begin{equation}
    \renewcommand\arraystretch{1.7}
    \vb{E} = \begin{pmatrix}
        A + B x \\
        - \frac{i \abs{\sin \alpha}}{\tan \alpha} \Big(A - \frac{g(\hat{\varepsilon}, \alpha)}{k_0 \sqrt{\varepsilon_\perp}} B + B x\Big) \\
        i \abs{\sin \alpha} \Big(A - \frac{\abs{\cot \alpha}}{k_0 \sqrt{\varepsilon_\perp}} B + B x\Big)
    \end{pmatrix} \mathrm{e}^{i q z + \lambda x}\label{eq:electric-field}
\end{equation}
where
\begin{align}
    g(\hat{\varepsilon}, \alpha) &= \abs{\cot \alpha} + \frac{4 \varepsilon_\perp}{\qty(\varepsilon_\parallel - \varepsilon_\perp) \abs{\sin 2\alpha}},\label{eq:g-func}\\
    q &= \abs{\cos \alpha}^{-1}\, k_0 \sqrt{\varepsilon_\perp}\label{eq:wavevector}\\
    \intertext{is the polariton wavevector,}
    \lambda &= \abs{\tan \alpha}\, k_0 \sqrt{\varepsilon_\perp}\label{eq:decay-constant}
\end{align}
is the localization constant, and \( k_0 = \omega / c \) is the vacuum wavevector.
Similarly in the upper crystal \( x > 0 \)
\begin{equation}
    \renewcommand\arraystretch{1.7}
    \vb{E} = \begin{pmatrix}
        A' + B' x \\
        \frac{i \abs{\sin \alpha'}}{\tan \alpha'} \Big(A' + \frac{g(\hat{\varepsilon}', \alpha')}{k_0 \sqrt{\varepsilon'_\perp}} B' + B' x\Big) \\
        - i \abs{\sin \alpha'} \Big(A' + \frac{\abs{\cot \alpha'}}{k_0 \sqrt{\varepsilon'_\perp}} B' + B' x\Big)
    \end{pmatrix} \mathrm{e}^{i q' z - \lambda' x},\label{eq:electric-field-up}
\end{equation}
but in Eqs.~\eqref{eq:g-func}--\eqref{eq:decay-constant} we need to add prime signs.
The magnetic field is calculated directly from Maxwell's equations.
The derivation of these equations is provided in Appendix \ref{app:singular-solution}.
The equality \( q^2 - \lambda^2 = k_0^2 \varepsilon_\perp \) is simply the dispersion relation.
Eq.~\eqref{eq:wavevector} impose a strict relation between surface polariton wavevector and its propagation direction relative to the optic axis.
The phase matching condition should be satisfied at the interface, hence \( q = q' \).
Which means that \( \alpha \) and \( \alpha' \) are not independent.
Coefficients \( A \), \( B \), \( A' \), and \( B' \) are determined from the boundary conditions for electromagnetic field.

We follow the notation used in \cite{Pancharatnam1958LightPropagation,Marchevskii1984SingularElectromagnetic} and call a surface polariton \textit{bisingular} if it is singular in both crystals.
In other studies \cite{Mackay2019DyakonovVoigt,Lakhtakia2020UnexceptionalDoubly,Lakhtakia2020ElectromagneticSurface} similar surface waves are called exceptional, doubly exceptional, or Dyakonov--Voigt waves.
It emphasizes a mathematically correct analogy with the exceptional points of Hamiltonians in quantum mechanics.
Bisingular surface waves appear to have been analyzed for the first time in \cite{Lakhtakia2020UnexceptionalDoubly}, at the interface of two biaxial crystals.
However, the dispersion equation was solved numerically because of its complexity.

Notably, the configuration under consideration is the only one that allows the propagation of singular surface polaritons.
If the optic axis is not parallel or orthogonal to the boundary, then the localization constant of extraordinary waves is a complex number for any propagation direction \cite{Fedorov1976TeoriyaGirotropiia}.
However, the localization constant of ordinary waves is always a real number.
Therefore, they cannot be equal, which is required for a singular solution.

\begin{figure*}
    \includegraphics[width=\linewidth]{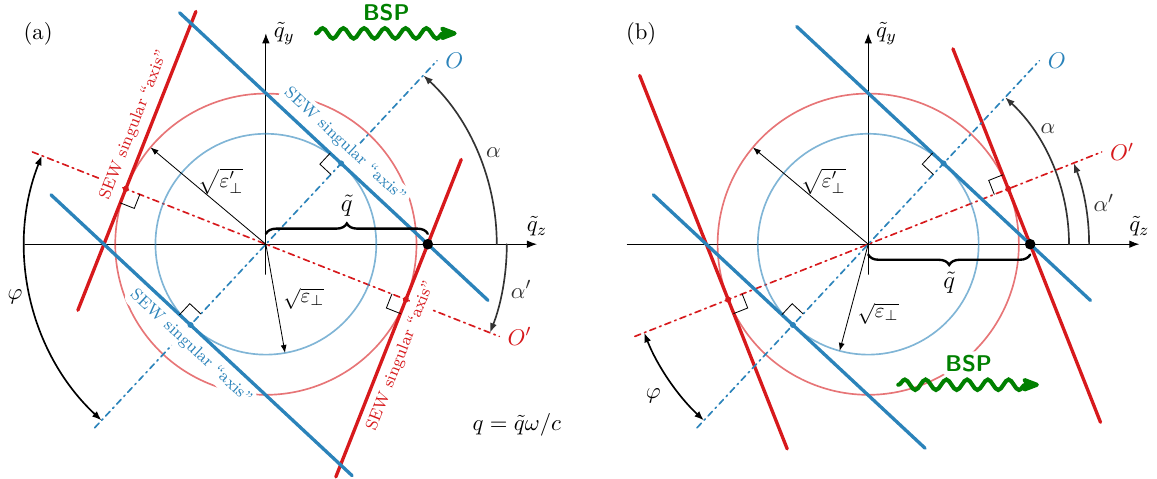}
    \caption{
        \label{fig:singular-axes} An example of the orientation of two crystals and the singular ``axes'' of surface waves in them:
        (a) \( \alpha > 0 \), \( \alpha' < 0 \), and \( \varphi \) satisfies \eqref{eq:dispersion-equation-diff}
        (b) \( \alpha > 0 \), \( \alpha' > 0 \), and \( \varphi \) satisfies \eqref{eq:dispersion-equation-same}.
        \( z \) axis always passes through the intersection point of the singular axes.
        It corresponds to the possible wavevector \( q = \tilde{q} k_0 \) \eqref{eq:wavevector} of the bisingular surface polariton (BSP).
    }
\end{figure*}

Without loss of generality we further assume that \( \varepsilon'_\perp \geqslant \varepsilon_\perp \) and \( 0 < \alpha < \pi/2 \).
The angle \( \alpha' \) can be either positive (Fig.~\Figref[(a)]{fig:singular-axes}) or negative (Fig.~\Figref[(b)]{fig:singular-axes}).

It seemed convenient to us to define the \textit{singular axes of surface waves} in order to visualize relation between angles.
Singular axes of surface waves are the set of wavevectors \( \vb{q} \) for which we are forced to choose a general solution for the surface wave in singular form.
In our case of a non-absorbing uniaxial medium, these are two lines
\begin{equation*}
    q_z \cos \alpha + q_y \sin \alpha = \pm \dfrac{\omega}{c} \sqrt{\varepsilon_\perp}
\end{equation*}
orthogonal to the optic axis and tangent to both light cones, with a punctured point of contact.
All intersection points of the singular axes of both media give possible \( \vb{q} \) for bisingular surface waves.
The definition of singular axes of surface waves can be extended to other types of anisotropic media (e.g. biaxial or absorbing media).
They will have a more complex shape than a two lines.
This is beyond the scope of this paper.

It is clear that for a given value \( q \) two angles \( \varphi \) are possible (Fig.~\ref{fig:singular-axes}).
Moreover, \( \alpha' > 0 \) for \( \varphi < \varphi_\star \) and \( \alpha' < 0 \) for \( \varphi > \varphi_\star \) where
\begin{equation}\label{eq:phi-star}
    \varphi_\star = \arccos \sqrt{\frac{\varepsilon_\perp}{\varepsilon'_\perp}}.
\end{equation}
It is easier to consider both cases separately, because Eqs.~\eqref{eq:electric-field}--\eqref{eq:electric-field-up} include the absolute values.
However, in some cases, for example identical crystals, only the case \( \alpha' < 0 \) is relevant.

All parameters of the bisingular surface polariton can be expressed in terms of \( \varphi \).
Using \( \alpha = \varphi + \alpha' \) we get
\begin{equation}\label{eq:wavevector-via-phi}
    q = \varepsilon'_\perp \sqrt{1 + \frac{(\cos\varphi - \cos\varphi_\star)^2}{\sin^2 \varphi}}.
\end{equation}
Angles \( \alpha \) and \( \alpha' \) is written as
\begin{align}
    \tan \alpha
    &= \dfrac{1-\cos \varphi_\star \cos\varphi}{\cos\varphi_\star \sin\varphi}\label{eq:tan-alpha},\\
    \tan \alpha'
    &= \dfrac{\cos \varphi - \cos \varphi_\star}{\sin\varphi}. \label{eq:tan-alpha-prime}
\end{align}
Now it is clear that \( \alpha' \) changes sign at \( \varphi = \varphi_\star \).
It follows from \eqref{eq:tan-alpha} that \( \tan \alpha \geqslant \tan \varphi_\star \) for any \( \varphi \).

A bisingular polariton does not exist at any angle \( \varphi \), because the electromagnetic fields must also satisfy boundary conditions at the interface.
The specific values of \( \varphi \) are determined by the dispersion equation.
Standard boundary conditions, for example continuity of \( E_y \), \( E_z \), \( D_x \), and \( H_y \) at the interface, lead to a system of linear equations on \( A, B, A', B' \).
The determinant of this system is the dispersion equation (see Appendix~\ref{app:dispersion-system} for details).
It gives additional equation involving wavevector \( q \) and angles \( \alpha \), \( \alpha' \).

After straightforward algebraic transformations one can obtain fairly simple  equations in terms of \( \varphi \).
In the case \( \alpha' < 0 \) it is written
\begin{gather}
    \begin{multlined}[b][0.8\linewidth]
        \Big(\textstyle\sqrt{\varepsilon'_\perp} + \sqrt{\varepsilon_\perp\vphantom{\varepsilon'}}\Big)^2 \Big(\varepsilon'_\perp + \varepsilon_\perp\Big) \sin^2 \dfrac{\varphi}{2} - \\
     \quad -\!2 \Big(\varepsilon'_\parallel - \varepsilon'_\perp\Big) \Big(\varepsilon_\parallel - \varepsilon_\perp\Big) \sin^2 \frac{\varphi}{2} \cos^4 \frac{\varphi}{2} + \\
     +\! \Big(\varepsilon'_\perp - \varepsilon_\perp\Big) \Big(\varepsilon'_\parallel - \varepsilon_\parallel\Big) \cos^2 \frac{\varphi}{2} = 0
    \end{multlined}\label{eq:dispersion-equation-diff}\\
    \intertext{subject to}
    \tan^2 \frac{\varphi}{2} > \frac{\sqrt{\varepsilon'_\perp} - \sqrt{\varepsilon_\perp\vphantom{\varepsilon'}}}{\sqrt{\varepsilon'_\perp} + \sqrt{\varepsilon_\perp\vphantom{\varepsilon'}}}.\label{eq:additional-condition}
\end{gather}
In the case \( \alpha' > 0 \) it is slightly different
\begin{gather}
    \begin{multlined}[b][0.8\linewidth]
    \Big(\textstyle\sqrt{\varepsilon'_\perp} - \sqrt{\varepsilon_\perp\vphantom{\varepsilon'}}\Big)^2 \Big(\varepsilon'_\perp + \varepsilon_\perp\Big) \cos^2 \dfrac{\varphi}{2} - \\
    -\! 2 \Big(\varepsilon'_\parallel - \varepsilon'_\perp\Big) \Big(\varepsilon_\parallel - \varepsilon_\perp\Big) \cos^2 \frac{\varphi}{2} \sin^4 \frac{\varphi}{2} + \\
    +\! \Big(\varepsilon'_\perp - \varepsilon_\perp\Big) \Big(\varepsilon'_\parallel - \varepsilon_\parallel\Big) \sin^2 \frac{\varphi}{2} = 0
    \end{multlined}\label{eq:dispersion-equation-same}\\
    \intertext{subject to}
    \label{eq:additional-condition-same}
    \tan^2 \frac{\varphi}{2} < \frac{\sqrt{\varepsilon'_\perp} - \sqrt{\varepsilon_\perp\vphantom{\varepsilon'}}}{\sqrt{\varepsilon'_\perp} + \sqrt{\varepsilon_\perp\vphantom{\varepsilon'}}}.
\end{gather}
Additional constraints are equivalent to choosing the sign of \( \alpha' \), or \( \varphi \lessgtr \varphi_\star \) \eqref{eq:phi-star}.
Both Eqs.~\eqref{eq:dispersion-equation-diff} and \eqref{eq:dispersion-equation-same} cover all possible combinations of uniaxial media.
The next section is devoted to the analysis of these equations and their roots.

\section{Analysis}
The main goal is to determine existence conditions of bisingular surface polaritons.
To do this, we need to determine when the roots of Eqs.~\eqref{eq:dispersion-equation-diff} and \eqref{eq:dispersion-equation-same} exist.
If there is at least one appropriate angle \( \varphi \), then the bisingular polariton exists.
Recall that the first necessary condition is \( \varepsilon'_\perp, \varepsilon_\perp > 0 \) which follows from \eqref{eq:wavevector}--\eqref{eq:decay-constant}.
Other conditions include anisotropy factors of each media
\begin{equation*}
    \eta = \dfrac{\varepsilon_\parallel}{\varepsilon_\perp} - 1, \qquad \eta' = \dfrac{\varepsilon'_\parallel}{\varepsilon'_\perp} - 1.
\end{equation*}
In general, they can be reduced to the requirement of ``strong'' anisotropy \( \abs{\eta} \gg 1 \), or \( \abs{\eta'} \gg 1 \).

Currently strong anisotropy can be achieved quite easily in various types of metamaterials \cite{Jahani2016AlldielectricMetamaterials}.
Another example, somewhat forgotten by researchers, is natural uniaxial minerals in the infrared and far-infrared range \cite{Korzeb2015CompendiumNatural,Galiffi2023ExtremeLight}.
Strong anisotropy is possible due to anisotropy of optical phonons, whose contribution to the permittivity is significant in this wavelength range.
It is more correct to call the solution a bisingular surface phonon polariton in this case.

Equations \eqref{eq:dispersion-equation-diff} and \eqref{eq:dispersion-equation-same} are inherently cubic in \( \cos^2 (\varphi / 2) \) or \( \sin^2 (\varphi / 2) \), and it makes analysis complicated in general case.
However, they are greatly simplified in several cases: \( \varepsilon'_\parallel = \varepsilon'_\perp \), \( \varepsilon_\parallel = \varepsilon_\perp \), \( \varepsilon_\perp' = \varepsilon_\perp \), or \( \varepsilon'_\parallel = \varepsilon_\parallel \).
The first two corresponds to the boundary of isotropic medium and anisotropic medium (Sec.~\ref{ref:subsec-weakly-isotropic-medium}).
The second two corresponds to identical or similar media (Sec.~\ref{ref:subsec-similar-media}).
Results for arbitrary media are given in Sec.~\ref{ref:subsec-arbitrary-media}.

\subsection{Isotropic medium or weakly anisotropic medium\label{ref:subsec-weakly-isotropic-medium}}
The first case \( \varepsilon'_\parallel = \varepsilon'_\perp \) corresponds to an isotropic medium in the upper half-space \( x > 0 \).
In a strict sense we should consider this only as limit \( \varepsilon'_\parallel \to \varepsilon'_\perp \).
In an isotropic medium, an evanescent wave never has the singular form \eqref{eq:electric-field}, and an isotropic medium has no optic axis.
However, it is still possible to obtain a solution.

Using \( \varepsilon'_\parallel = \varepsilon'_\perp = \varepsilon_i \) we get
\begin{align}
    \tan^2 \dfrac{\varphi_{i,1}}{2} &= \dfrac{\qty(\varepsilon_\parallel - \varepsilon_i)\qty(\sqrt{\varepsilon_i} - \textstyle\sqrt{\varepsilon_\perp})}{\qty(\varepsilon_\perp + \varepsilon_i) \qty(\sqrt{\varepsilon_i} + \textstyle\sqrt{\varepsilon_\perp})} \nonumber\\
    &= \dfrac{\eta\cos^2\varphi_\star-\sin^2\varphi_\star}{1+\cos^2\varphi_\star}\, \tan^2 \dfrac{\varphi_\star}{2} \label{eq:phi-isotropic-diff}
\end{align}
from \eqref{eq:dispersion-equation-diff} and
\begin{align}
    \cot^2 \dfrac{\varphi_{i,2}}{2} &= \dfrac{\qty(\varepsilon_\parallel - \varepsilon_i)\qty(\sqrt{\varepsilon_i} + \textstyle\sqrt{\varepsilon_\perp})}{\qty(\varepsilon_\perp + \varepsilon_i) \qty(\sqrt{\varepsilon_i} - \textstyle\sqrt{\varepsilon_\perp})} \nonumber\\
    &= \dfrac{\eta\cos^2\varphi_\star - \sin^2\varphi_\star}{1 + \cos^2\varphi_\star}\, \cot^2 \dfrac{\varphi_\star}{2} \label{eq:phi-isotropic-same}
\end{align}
from \eqref{eq:dispersion-equation-same}.
The solutions are obviously different, but this should not be confusing.
In this case \( \varphi \) has no physical interpretation because \( \alpha' \) cannot be defined in isotropic medium.
However, angle \( \alpha \) has a clear meaning.
Using tangent half-angle formulae in \eqref{eq:tan-alpha}, we find that
\begin{align}
    \tan^2 \alpha &= \dfrac{\qty(\varepsilon_i - \varepsilon_\perp) \qty(\varepsilon_\parallel + \varepsilon_\perp)^2}{4 \varepsilon_\perp \qty(\varepsilon_\perp + \varepsilon_i)\qty(\varepsilon_\parallel - \varepsilon_i)} \nonumber\\
    &= \dfrac{(\eta + 2)^2 \cos^2 \varphi_\star }{4(1 + \cos^2 \varphi_\star) \qty(\eta \cot^2 \varphi_\star - 1)}\label{eq:alpha-isotropic}
\end{align}
is the same for both solutions \eqref{eq:phi-isotropic-diff}, \eqref{eq:phi-isotropic-same}.
The corresponding angles \( \alpha' \) differ only in sign
\begin{align}
    \tan \alpha'_{1,2} &= \mp \frac{\varepsilon_\parallel - \varepsilon_\perp - 2 \varepsilon_i}{2 \sqrt{\varepsilon_i \qty(\varepsilon_i + \varepsilon_\perp)}} \sqrt{\frac{\varepsilon_i - \varepsilon_\perp}{\varepsilon_\parallel - \varepsilon_i}}\nonumber\\
    &= \mp \dfrac{\eta \cos^2 \varphi_\star - 2}{2 \sqrt{\qty(1 + \cos^2 \varphi_\star) \qty(\eta \cot^2 \varphi_\star - 1)}} \label{eq:tan-alpha-prime-isotropic}.
\end{align}
Sign ``\( - \)'' refers to solution \eqref{eq:phi-isotropic-diff}, and sign ``\( + \)'' to \eqref{eq:phi-isotropic-same}.

Obtained expression~\eqref{eq:alpha-isotropic} is the same as given in \cite{Marchevskii1984SingularElectromagnetic,Mackay2019DyakonovVoigt,Golenitskii2024AnisotropicSurface} subject to \( \varepsilon_i > \varepsilon_\perp \).
Constraints \eqref{eq:additional-condition} and \eqref{eq:additional-condition-same} yield the same existence condition \( \varepsilon_\parallel > \varepsilon_\perp + 2 \varepsilon_i \).
This can also be seen from \eqref{eq:tan-alpha-prime-isotropic}.
Therefore, it is required that at least \( \varepsilon_\parallel > 3 \varepsilon_\perp \), or equivalently \( \eta > 2 \).
This limit corresponds to dielectric media because all dielectric constants should be positive.

Weak anisotropy \( \varepsilon'_\parallel = \varepsilon'_\perp (1 + \eta') \), \( \eta' \ll 1 \) can be taken into account to get a ``fair'' bisingular surface polarion.
An approximate solution \( \varphi_{1,2} \) can be found as an asymptotic series in the small parameter \( \eta' \).
We have calculated first-order correction terms
\begin{align}
    \frac{\tan^2 \qty({\varphi_1}/{2})}{\tan^2 \qty({\varphi_{i,1}}/{2})} &\approx 1 +  \frac{\eta' F}{\qty(2 + \eta \cos\varphi_\star \tan^2\qty({\varphi_\star}/{2}))^2},\label{eq:phi-weak-anisotropy-1}\\
    \frac{\tan^2 \qty({\varphi_2}/{2})}{\tan^2 \qty({\varphi_{i,2}}/{2})} &\approx 1 - \frac{\eta' F}{\qty(2 - \eta \cos\varphi_\star \cot^2 \qty({\varphi_\star}/{2}))^2}\label{eq:phi-weak-anisotropy-2}
    \intertext{where}
    F &= \frac{\qty(\eta \cos^2 \varphi_\star - 2) (\eta + 2)}{\eta \cos^2 \varphi_\star - \sin^2 \varphi_\star}.\label{eq:a-weak-anisotropy}
\end{align}
When analyzing the conditions of existence, it is convenient to rewrite \( \varepsilon_\parallel > \varepsilon_\perp + 2 \varepsilon'_\perp \) as \( \eta \cos^2 \varphi_\star > 2 \).
It can be shown that the existence condition of a bisingular polariton does not change regardless of the sign of \( \eta' \).
Thus the first correction term depends only on the sign of \( \eta' \) because \( F > 0 \) is always true.
It should also be noted that the denominator in \eqref{eq:phi-weak-anisotropy-2} never vanishes.

In the second case, the other medium is considered to be isotropic \( \varepsilon_\| = \varepsilon_\perp = \varepsilon_i \).
Now the only angle \( \alpha' \) has physical meaning.
Carrying out similar calculations, we get \( \tan^2\alpha' \).
It turns out to be the same as \eqref{eq:alpha-isotropic} if we make substitutions \( \varepsilon_\perp \) for \( \varepsilon'_\perp \), and \( \varepsilon_\parallel \) for \( \varepsilon'_\parallel \).
In fact, it is possible to verify that the stronger statement is true.
Eqs.~\eqref{eq:phi-isotropic-diff}, \eqref{eq:phi-isotropic-same}, and \eqref{eq:phi-weak-anisotropy-1}--\eqref{eq:a-weak-anisotropy} can be applied to this case too if we simultaneously replace \( \eta' \to - \eta \cos^2 \varphi_\star \) and \( \eta \to - \eta' / \cos^2 \varphi_\star \).
In this case existence condition for bisingular polariton is \( \varepsilon'_\parallel < - \varepsilon'_\perp < 0 \) and covers another case mentioned in \cite{Marchevskii1984SingularElectromagnetic}.
It corresponds to polariton at the boundary of a hyperbolic uniaxial medium and isotropic dielectric (e.g. air).

Let us sum up this part.
Bisingular surface polariton exists at the boundary of two uniaxial media, one of which is weakly anisotropic, in two cases.

In the first case both media are dielectrics \( \varepsilon_\parallel > 0 \), \( \varepsilon'_\parallel \approx \varepsilon'_\perp > \varepsilon_\perp > 0 \).
The existence condition requires \( \varepsilon_\parallel > \varepsilon_\perp + 2 \varepsilon'_\perp \).
If it is fulfilled, then there are two angles between optic axes \eqref{eq:phi-weak-anisotropy-1}, \eqref{eq:phi-weak-anisotropy-2} for which a bisingular surface polariton exists.

In the second case, one of the media should be hyperbolic \( \varepsilon'_\parallel < -\varepsilon'_\perp < 0 \) and the other dielectric \( \varepsilon_\parallel \approx \varepsilon_\perp > 0 \).
Similar to the previous case, there are two angles for which a bisingular surface polariton exists.
They are given by the same equations \eqref{eq:phi-weak-anisotropy-1}, \eqref{eq:phi-weak-anisotropy-2}, but with the replacement \( \eta' \to - \eta \cos^2 \varphi_\star \) and \( \eta \to - \eta' / \cos^2 \varphi_\star \).

In both cases, it turns out that the bisingular polariton propagates under conditions close to those described in \cite{Marchevskii1984SingularElectromagnetic} for a singular surface polariton.
Weak anisotropy does not affect the existence condition, even for corner cases (\( \varepsilon'_\perp \approx \varepsilon_\perp \), or \( \varepsilon_\parallel \approx \varepsilon_\perp + 2 \varepsilon'_\perp \), or \( \varepsilon'_\parallel \approx - \varepsilon'_\perp \)).
Our preliminary estimates show that for large \( \eta \) (or \( \abs{\eta'} \)) and \( \varepsilon'_\perp \approx \varepsilon_\perp \) requires a more careful analysis.
This is due to the possible appearance of new solutions discussed in the next section.

%

\subsection{Similar media\label{ref:subsec-similar-media}}
The second type of solutions pertains to the case of media in which at least some components of the dielectric tensors coincide, specifically \( \varepsilon'_\perp = \varepsilon_\perp \) or \( \varepsilon'_\parallel = \varepsilon_\parallel \).
Due to this, the last term in Eqs.~\eqref{eq:dispersion-equation-diff}, \eqref{eq:dispersion-equation-same} vanishes.
To the best of the authors' knowledge, the analytical solutions obtained in this section have not been previously described.

Let us first consider the case of identical media.
It is known that if these are dielectric positive uniaxial crystals (\( \varepsilon_\parallel > \varepsilon_\perp \)), then the existence of Dyakonov surface waves is predicted \cite{Averkiev1990ElectromagneticWaves}.
They propagate in a range of angles around the bisector between the optic axes.
However, the authors omitted the special cases of singular and bisingular surface waves which also should exist in a highly anisotropic crystals.
The case of singular surface polaritons (the singular solution \eqref{eq:electric-field} in one crystal, and the sum of an ordinary and extraordinary waves in the second) is beyond the scope of this work.
Therefore, let us dwell only on the bisingular surface polariton.
From \eqref{eq:dispersion-equation-diff} and \eqref{eq:tan-alpha} we get solution
\begin{equation}\label{eq:phi-identical-crystals}
    \cos^2 \dfrac{\varphi_\text{AD}}{2} = \cos^2 \alpha = \dfrac{2}{\abs{\eta}}
\end{equation}
that requires \( \abs{\eta} > 2 \) for the existence of angle \( \varphi_\text{AD} \).
This means that the bisingular polariton exists not only for dielectric crystals, but also for hyperbolic.
In terms of dielectric constants \( \varepsilon_\parallel > 3 \varepsilon_\perp \), or \( \varepsilon_\parallel < -\varepsilon_\perp \).
Bisingular surface polariton in this case propagates exactly along the bisector of the angle formed by the optic axes, so \( \alpha = \varphi / 2 \) and \( \alpha' = - \varphi / 2 \).
This can also be seen clearly by plotting diagrams (\Figref[]{fig:singular-axes}) for \( \varepsilon'_\perp = \varepsilon_\perp \).

Let us note that Eq.~\eqref{eq:phi-identical-crystals} can also be obtained from the Dyakonov surface wave dispersion equation \cite[Eq.~(6) in][]{Averkiev1990ElectromagneticWaves}, derived without considering singular solutions.
To do this we need to set the localization constants of ordinary and extraordinary waves equal to each other \( k_0 = k_1 = k_2 \).
After algebraic transformations we get Eq.~\eqref{eq:phi-identical-crystals}.
However, the expressions for electromagnetic fields in \cite{Averkiev1990ElectromagneticWaves} are not applicable.
It can be verified that ordinary and extraordinary wave has the same polarization.
That is, they correspond to the same solution and do not form a complete basis of eigenwaves.

The case of identical media can easily be extended to \( \varepsilon'_\parallel \neq \varepsilon_\parallel \).
Keeping \( \varepsilon'_\perp = \varepsilon_\perp \) in mind, we get
\begin{equation}\label{eq:phi-similar-perps}
    \cos^4 \dfrac{\varphi_{\perp}}{2} = \cos^4 \alpha_\perp = \dfrac{4}{\eta \eta'} = \dfrac{4 \varepsilon^2_\perp}{\qty(\varepsilon_\parallel - \varepsilon_\perp)\big(\varepsilon'_\parallel - \varepsilon_\perp\big)}.
\end{equation}
Such an angle \( \varphi \) exists only if \( \eta \eta' > 4 \), that is, they are of the same sign.
There are only three possible combinations of media types: both media are positive dielectrics, both media are hyperbolic, one is hyperbolic and other is a negative dielectric.
The last combination is interesting because it combines different types of media.

As in the previous section, an approximate solution by the small parameter \( (\varepsilon'_\perp - \varepsilon_\perp) / \varepsilon_\perp \ll 1 \), or \( \varphi_\star \ll 1 \) equivalently, can be found.
After expanding \eqref{eq:dispersion-equation-diff} up to a term of order \( \varphi_\star^2 \), we obtain
\begin{equation}\label{eq:similar-perp-approx-equation}
    \cos^4 \dfrac{\varphi}{2} \approx \dfrac{4}{\eta\eta'}\qty(1 + \dfrac{\eta' - \eta}{8} \cot^2\dfrac{\varphi}{2}\, \varphi_\star^2).
\end{equation}
It can have up to two solutions.
One is close to \eqref{eq:phi-similar-perps} in some case, as expected.
Another solution is \( \varphi \propto \varphi_\star \) and disappears at \( \varphi_\star = 0 \).
It is only possible because of \( \cot^2 (\varphi / 2) \) on the right side.

The approximate solution to \eqref{eq:phi-similar-perps} is
\begin{equation}
    \cos^4 \dfrac{\varphi_{\perp,1}}{2} \approx \dfrac{4}{\eta\eta'}\qty(1 + \dfrac{\eta' - \eta}{\sqrt{\eta\eta'} - 2} \dfrac{\varphi_\star^2}{4})
\end{equation}
and it only works if \( (\eta' - \eta) \varphi_\star^2 / \varphi^2_\perp \ll 1 \) is satisfied.
So it is not applicable for \( \varphi_\perp \) close to \( 0 \).
Both solution are mixed up, and it is not possible to build a simple perturbative series.
An additional solution from equation \eqref{eq:dispersion-equation-same} may also appear.

All of these solutions may overlap with the approximate solutions \eqref{eq:phi-weak-anisotropy-1}, \eqref{eq:phi-weak-anisotropy-2} from the previous section if \( \varphi \propto \varphi_\star \).
It seems to us that it is more productive to solve the equations exactly in these cases (see Sec.~\ref{ref:subsec-arbitrary-media}).

The last combination that allows a simple solution is \( \varepsilon'_\parallel = \varepsilon_\parallel \), and \( \varepsilon'_\perp \neq \varepsilon_\perp \) is possible.
Parameters \( \eta \), \( \eta' \), and \( \varphi_\star \) are no longer independent.
The connection between them is given
\begin{equation*}
    \varepsilon'_\parallel - \varepsilon_\parallel = \varepsilon'_\perp \qty(1 + \eta' - (1 + \eta) \cos^2 \varphi_\star) = 0
\end{equation*}

In contrast to the previous case, there are two solutions possible.
The first is
\begin{align}
    \cos^4 \frac{\varphi_{\parallel,1}}{2} &= \frac{4 + 2\tan^2 \varphi_\star}{\eta \eta'} \cos^4 \frac{\varphi_\star}{2} \nonumber\\
    &= \frac{\qty(\sqrt{\varepsilon'_\perp} + \sqrt{\varepsilon_\perp})^2 (\varepsilon'_\perp + \varepsilon_\perp)}{2 ( \varepsilon_\parallel - \varepsilon_\perp ) ( \varepsilon_\parallel - \varepsilon'_\perp )}\label{eq:phi-similar-parallel-1}
\end{align}
and follows from Eq.~\eqref{eq:dispersion-equation-diff}.
The second is
\begin{align}
    \sin^4 \frac{\varphi_{\parallel,2}}{2} &= \frac{4 + 2\tan^2 \varphi_\star}{\eta \eta'} \sin^4 \frac{\varphi_\star}{2} \nonumber\\
    &= \frac{\qty(\sqrt{\varepsilon'_\perp} - \sqrt{\varepsilon_\perp})^2 (\varepsilon'_\perp + \varepsilon_\perp)}{2 ( \varepsilon_\parallel - \varepsilon_\perp ) ( \varepsilon_\parallel - \varepsilon'_\perp )}\label{eq:phi-similar-parallel-2}
\end{align}
and follows from Eq.~\eqref{eq:dispersion-equation-same}.
Constraints on \( \varphi \) \eqref{eq:additional-condition} and \eqref{eq:additional-condition-same}, respectively, lead to the same condition \begin{equation}\label{eq:condition-similar-parallel}
    \eta\eta' > 4 + 2\tan^2 \varphi_\star.
\end{equation}
Therefore, both solutions either exist simultaneously or do not.
As \( \varepsilon'_\perp \to \varepsilon_\perp \), solution \( \varphi_{\parallel,2} \to 0 \) and disappears, but \( \varphi_{\parallel,2} \to \varphi_\text{AD} \) \eqref{eq:phi-identical-crystals} at the same time.

It is clear that the bisingular polariton in this case propagates in a direction different from the bisector of the angle between the optic axes.
Due to the form of Eqs.~\eqref{eq:phi-similar-parallel-1}, \eqref{eq:phi-similar-parallel-2}, it is more convenient to find \( \alpha' \) \eqref{eq:tan-alpha-prime}
\begin{align}
    \tan \alpha'_{\parallel,1} = \qty(\sqrt{\dfrac{4 + 2 \tan^2 \varphi_\star }{\eta \eta'}} - 1) \dfrac{2 \cos^2 \frac{\varphi_\star}{2}}{\sin \varphi_{\parallel,1}}, \\
    \tan \alpha'_{\parallel,2} = \qty(1 - \sqrt{\dfrac{4 + 2 \tan^2 \varphi_\star }{\eta \eta'}}) \dfrac{2 \sin^2 \frac{\varphi_\star}{2}}{\sin \varphi_{\parallel,2}}.
\end{align}
We make sure that the correct sign of \( \alpha' \) will only be if \eqref{eq:condition-similar-parallel} is met.

As in the previous sections, let us try to find approximate solution for \( \varepsilon'_\parallel - \varepsilon_\parallel \ll \varepsilon_\perp \).
A rough estimate can be obtained by considering that the last terms \( \propto (\varepsilon'_\parallel - \varepsilon_\parallel) \) in \eqref{eq:dispersion-equation-diff}, \eqref{eq:dispersion-equation-same} give only a small correction.
Without an exact series expansion we get
\begin{align*}
    \dfrac{\cos^4 \qty({\varphi}/{2})}{\cos^4 \qty({\theta_{\parallel,1}}/{2})} \approx 1 + \dfrac{2 \tan^2 \varphi_\star}{\eta \eta' \sin^2 \theta_{\parallel,1} } \dfrac{\varepsilon'_\parallel - \varepsilon_\parallel}{\varepsilon'_\perp},\\
    \intertext{from \eqref{eq:dispersion-equation-diff} and}
    \dfrac{\sin^4 \qty({\varphi}/{2})}{\sin^4 \qty({\theta_{\parallel,2}}/{2})} \approx 1 + \dfrac{2 \tan^2 \varphi_\star}{\eta \eta' \sin^2 \theta_{\parallel,2} } \dfrac{\varepsilon'_\parallel - \varepsilon_\parallel}{\varepsilon'_\perp}
\end{align*}
from \eqref{eq:dispersion-equation-same}.
Angles \( \theta_{\parallel,1} \) and \( \theta_{\parallel,2} \) are defined by the first equalities in \eqref{eq:phi-similar-parallel-1} and \eqref{eq:phi-similar-parallel-2}, respectively, but with exact values \( \eta \), \( \eta' \).
It can be seen that problems in these approximations arise if ``zeroth'' order solution \( \theta_{\parallel,j} \) close to \( 0 \) or \( \pi \), where \( j = 1, 2 \).
It become even worse if difference between \( \varepsilon'_\perp \) and \( \varepsilon_\perp \) is large due to \( \tan^2 \varphi_\star \).

\subsection{Arbitrary media\label{ref:subsec-arbitrary-media}}
It is clear that in the general case an analytical solution of Eqs.~\eqref{eq:dispersion-equation-diff}, \eqref{eq:dispersion-equation-same} can also be written out.
However, taking into account restrictions \eqref{eq:additional-condition} and \eqref{eq:additional-condition-same}, the question of the existence of acceptable roots and their number seems more interesting.

Equations \eqref{eq:dispersion-equation-diff} and \eqref{eq:dispersion-equation-same}, due to their similarity, can be reduced to the same cubic equation.
It significantly simplifies further analysis.
Let us denote \begin{equation}
    t_* = \tan^2 (\varphi_\star / 2) = \dfrac{\sqrt{\varepsilon'_\perp} - \sqrt{\varepsilon_\perp}}{\sqrt{\varepsilon'_\perp} + \sqrt{\varepsilon_\perp}}
\end{equation}
since it occurs frequently.
Possible values are bounded \( 0 \leqslant t_\star < 1 \).
Making the substitution
\begin{equation*}
    t = \dfrac{1 + \tan^2 (\varphi / 2)}{1 + t_\star} = \dfrac{\cos^2 (\varphi_\star / 2)}{\cos^2 (\varphi / 2)}
\end{equation*}
in \eqref{eq:dispersion-equation-diff} and
\begin{equation*}
    t = \dfrac{1 + \cot^2 (\varphi / 2)}{1 + t_\star^{-1}} = \dfrac{\sin^2 (\varphi_\star / 2)}{\sin^2 (\varphi / 2)}
\end{equation*}
in \eqref{eq:dispersion-equation-same} reduces both equations to
\begin{equation}\label{eq:t-cubic-equation}
     P(\tau, t) \equiv \qty(t^2 - \Delta_2) \qty(t - \dfrac{1 + \Delta_1 \tau}{1 + \tau}) - \dfrac{\Delta_1 \Delta_2 \tau}{1 + \tau} = 0
\end{equation}
subject to \( t > 1 \), where \( \tau = t_\star \) for \eqref{eq:dispersion-equation-diff} and \( \tau = t_\star^{-1} \) for \eqref{eq:dispersion-equation-same}.
Parameters \( \Delta_1 \) and \( \Delta_2 \) are defined as
\begin{align*}
    \Delta_1 &= \dfrac{\varepsilon_\parallel - \varepsilon'_\parallel}{\varepsilon'_\perp + \varepsilon_\perp} = \dfrac{\eta - \eta'}{2} - \dfrac{\eta + \eta' + 2}{\tau + \tau^{-1}}, \\
    \Delta_2 &= \dfrac{(\varepsilon_\parallel - \varepsilon_\perp)(\varepsilon'_\parallel - \varepsilon'_\perp)}{2 \varepsilon'_\perp (\varepsilon'_\perp + \varepsilon_\perp)} = \dfrac{\eta\eta'}{4}\qty(1 - \dfrac{2}{\tau + \tau^{-1}})
\end{align*}
and are the same at \( \tau = t_\star \) and \( \tau = t_\star^{-1} \).

Three cases \( \Delta_1 = 0 \), \( \Delta_2 = 0 \), and \( t_\star = 0 \) were analyzed in previous sections.
We further assume that all of them are nonzero.
Roots of \( P(\tau, t) \) for \( \tau = t_\star \) and \( \tau = t_\star^{-1} \) correspond uniquely to different solutions (see~Fig.~\Figref{fig:singular-axes}).

Some restrictions on dielectric permittivities for which a bisingular polariton exists can be obtained without solving the equation.
It is known that any cubic polynomial always has one real root.
Therefore, if the value \( P(\tau, t) < 0 \) at \( t = 1 \), then there necessarily exists at least one root satisfying \( t > 1 \).
That is, if condition
\begin{equation}\label{eq:root-existence-condition}
    1 - \Delta_1 - \Delta_2 < 0
\end{equation}
is satisfied, then the bisingular surface polariton exists.
Rewriting in terms of dielectric permittivities gives
\begin{equation}\label{eq:one-root-condition}
    \qty(\varepsilon_\parallel - \varepsilon_\perp - 2 \varepsilon'_\perp) \qty(\varepsilon'_\parallel + \varepsilon'_\perp) > 0.
\end{equation}
It is quite broad and covers a wide range of media including dielectrics and hyperbolic media.
There are also parameter domains where condition \eqref{eq:root-existence-condition} is not satisfied, but there is a suitable root(s).

The question of the number of roots is more complicated.
First, note that \eqref{eq:root-existence-condition} does not depend on \( \tau \).
Secondly, if \( P(t_\star, t) \) and \( P(t_\star^{-1}, t) \) intersect, it happens at \( t = \pm \sqrt{\Delta_0 / (1 - \Delta_1)} \).
It follows from this that roots of \( P(t_\star, t) \) and \( P(t_\star^{-1}, t) \) appear simultaneously, in pairs.
Multiple roots are also count.
The total number of roots of \( P(t_\star, t) \) and \( P(t_\star^{-1}, t) \) is always even, and the number of different roots can be from two to six.
It is clear that there are always two roots if \eqref{eq:one-root-condition} is met.
Six roots are possible if \( t_\star \approx 1 \) (\( \varepsilon'_\perp \gg \varepsilon_\perp \)) because both polynomials are the same \( P(t_\star, t) \approx P(t_\star^{-1}, t) \) as \( t_\star \to 1 \).

To confirm our analysis we have carried out a numerical study of \eqref{eq:t-cubic-equation}.
In most cases \( P(t_\star, t) \) and \( P(t_\star^{-1}, t) \) simultaneously have only two solutions in total.
However, it is possible to find such parameters \( t_\star, \Delta_1, \Delta_2 \) when all three roots of \( P(t_\star, t) \) can satisfy the constraint \( t > 1 \).
But they require an extremely high degree of anisotropy and are very likely not achievable, for example \( t_\star = 0.25, \Delta_1 = 32, \Delta_2 = -17.9 \).

Analysis of the zeros of the derivative \( dP(\tau, t)/dt \) together with the value \( P(\tau, 1) \) shows that there are no roots \( t > 1 \) if both \( \Delta_1 < 0 \) and \( \Delta_2 < 0 \).
It is rewritten in terms of dielectric constants as \( \varepsilon'_\parallel > \varepsilon_\parallel \) and \( (\varepsilon_\parallel - \varepsilon_\perp)(\varepsilon'_\parallel - \varepsilon'_\perp) < 0 \).
Bisingular surface polariton does not exist for such pair of media.

\section{Conclusions}
Special surface polaritons propagating along the boundary of two arbitrary uniaxial media and called bisingular have been investigated in the study.
Bisingular means that field profile of the polariton is of the form \( (\vb{f}_1 + \vb{f}_2 x) \exp(-\lambda \abs{x}) \) in both media.
In essence, a solution in this form arises when the localization constants of eigenwaves in an anisotropic medium are equal.
This makes it possible to greatly simplify the analysis of the dispersion equation and also solve it analytically.
The analytical solution of a singular surface polariton at the boundary of an isotropic and uniaxial medium has been extended to the case of two uniaxial media.

It is shown that a bisingular surface polariton exists only when the optic axes of both media are parallel to the boundary.
All characteristics (wavevector, propagation direction, localization constants, field distributions) of the polariton, if it exists, can be expressed in terms of the single parameter, the angle between the optic axes.
The dispersion equation reduces to two similar cubic equations at this angle with simple constraints.
We have analyzed them in detail.
Depending on dielectric permittivities of media, they either have no solutions or have from two to six solutions.
Thus, up to six mutual orientations of the two given media are possible, allowing propagation of the bisingular polariton.
Some general sufficient conditions for the existence and non-existence of a bisingular polariton have also been found.
The analysis is simplified even further in some cases: one of the media is slightly anisotropic; both media are ``almost'' the same.
In these cases simple approximate solutions have been obtained.

Experimental observation of singular surface polaritons of any kind, including bisingular, is probably a difficult task.
The problems are not related to finding suitable media, they can be found, or to taking into account absorption.
The main problem is that singular polaritons do not propagate over a range of angles, but along a specific direction.
However, it is clear that near these directions there should also exist non-singular surface polaritons, such as Dyakonov surface waves.
Their existence conditions are more strict than for non-singular polaritons.
The existence of singular polaritons suggests near which directions in the boundary plane surface waves should be sought.
On the other hand, the problem of propagation of surface waves along the boundary of arbitrary anisotropic media is hard to be solved in the general case.
Particular analytical solutions, bisingular polaritons in our case, help to better understand in which direction to move further.

\section*{Acknowledgments}
KYuG acknowledge financial support from the ITMO--MIPT--Skoltech Clover Program.

\appendix
\section{Singular solution\label{app:singular-solution}}
The appearance of singular waves \eqref{eq:electric-field} was explained in \cite{Khapalyuk1962TheoryCircular,Fedorov1976TeoriyaGirotropiia,Marchevskii1984SingularElectromagnetic,Mackay2019DyakonovVoigt,Lakhtakia2020UnexceptionalDoubly}.
Here we want to clearly demonstrate the possibility of the existence of singular solutions and the analogy with exceptional points in other systems.

We seek Maxwell's equations in anisotropic media
\begin{align*}
    \curl \qty(\curl \vb{E}) &= -\dfrac{1}{c^2} \pdv[2]{\hat{\varepsilon} \vb{E}}{t} \\
    \div \qty(\hat{\varepsilon} \vb{E}) &= 0
\end{align*}
in the form of a wave \( \vb{E}(\vb{r}) = \vb{E}(x) \exp(i q z - i \omega t) \) propagating along \( z \) axis.
We keep in mind that \( \vb{E}(x) \to 0 \) as \( \abs{x} \to \infty \) is required for wave localized at the interface between media.
The magnetic field can be directly calculated \( \vb{H} = \qty(c / i \omega) \curl \vb{E} \).

The chosen coordinate system does not coincide with the principal axes of dielectric tensor \( \hat{\varepsilon} \) but optic axis is parallel to the interface plane \( (yz) \).
Let \( \alpha > 0 \) to be the angle corresponding to counterclockwise rotation of the media around the \( x \) axis.

Making the substitution \( \vb{E}(\vb{r}) \) reduces Maxwell's equations to the system of ordinary differential equations for three unknown functions \( \vb{E}(x) = (E_x, E_y, E_z) \).
One of the components is expressed explicitly through the other two
\begin{equation*}
    E_z = - \dfrac{\varepsilon_\perp}{i q \varepsilon_{zz}} \dv{E_x}{x} - \dfrac{\varepsilon_{zy}}{\varepsilon_{zz}} E_y.
\end{equation*}
The remaining two equations are coupled
\begin{align*}
    &\varepsilon_\perp \dv[2]{E_x}{x} + \qty(k_0^2 \varepsilon_\perp - q^2) \varepsilon_{zz} E_x + i q \varepsilon_{zy} \dv{E_y}{x} = 0,\\
    &\varepsilon_{zz} \dv[2]{E_y}{x} + \qty(k_0^2 \varepsilon_\perp \varepsilon_\parallel - q^2 \varepsilon_{zz}) E_y - \dfrac{\varepsilon_{yz} \varepsilon_\perp k_0^2}{i q} \dv{E_x}{x} = 0,
\end{align*}
where
\begin{align*}
    \varepsilon_{zz} &= \varepsilon_\parallel \cos^2 \alpha + \varepsilon_\perp \sin^2 \alpha, \\
    \varepsilon_{zy} &= \varepsilon_{yz} = {(\varepsilon_\parallel - \varepsilon_\perp)} \cos \alpha \sin \alpha,
\end{align*}
and \( k_0 = \omega / c \).
Now it is clear that the presence of \( \varepsilon_{yz} \) leads to coupling of waves of two polarizations.
If the medium is isotropic \( \varepsilon_\parallel - \varepsilon_\perp = 0 \), the propagation is along the optic axis or perpendicular to it then \( \varepsilon_{yz} = 0 \), and waves are decoupled.
The waves in these cases are very well-known and are usually called TE (\( E_y \)) and TM (\( E_x \)).

The system can be looked at from a different perspective.
Let \( E_x \) and \( E_y \) be the amplitudes of two oscillators, and \( x \) be a kind of time variable.
Then \( \varepsilon_{yz} \) and \( \varepsilon_{zy} \) describe a dissipative coupling between two oscillators.
It is known that in such a system the existence of exceptional points is possible.

Seeking the solution of the system in an exponential form \( E_x, E_y \propto \exp(\lambda x) \) leads to characteristic equation
\begin{multline}\label{eq:characteristic-equation}
    \qty(\dfrac{\varepsilon_\perp}{\varepsilon_{zz}} \lambda^2 + k_0^2 \varepsilon_\perp - q^2) \qty( \lambda^2 + k_0^2 \dfrac{\varepsilon_\perp \varepsilon_\parallel}{\varepsilon_{zz}} - q^2) + \\
    + \lambda^2 \dfrac{\varepsilon_\perp \varepsilon_{yz}^2}{\varepsilon_{zz}} k_0^2 = 0.
\end{multline}
It has four distinct roots \( \lambda \) if determinant is not zero.
Two pair of roots, differing in sign, correspond to the ordinary and extraordinary waves.
The analysis of surface waves at the boundary of two identical uniaxial media in this non-degenerate case was carried out in \cite{Averkiev1990ElectromagneticWaves}.

In the case that is interesting to us, the discriminant of \eqref{eq:characteristic-equation} should be equal to zero.
This possible if and only if
\begin{equation*}
    q^2 = \dfrac{\varepsilon_\perp k_0^2}{\cos^2 \alpha}.
\end{equation*}
Then Eq.~\eqref{eq:characteristic-equation} has two roots of multiplicity two
\begin{equation*}
    \lambda = \pm k_0 \sqrt{\varepsilon_\perp} \abs{\tan \alpha}.
\end{equation*}
If \( \alpha \neq 0 \) then equations remain coupled, and other linearly independent solutions we have to take in the form \( x \exp(\pm \lambda x) \).
It is called singular solution.

By choosing the correct signs for \( \lambda \) in each media and substituting \( E_x = (A + B x) \exp(\pm\lambda x) \), we obtain expressions \eqref{eq:electric-field}--\eqref{eq:electric-field-up} for electric field.

\section{Eigenwaves coefficients\label{app:dispersion-system}}
Electric \eqref{eq:electric-field}--\eqref{eq:electric-field-up} and magnetic field should satisfy standard boundary conditions for Maxwell's equations at \( x = 0 \).
We can choose any four components whose boundary conditions lead to four linearly independent equations on \( A, B, A', B' \)
\begin{widetext}
    \begin{align*}
    D_x:\quad &\varepsilon'_\perp A' = \varepsilon_\perp A, \\
    E_y:\quad &- i \dfrac{\abs{\sin \alpha}}{\tan \alpha} \qty(A - \dfrac{g(\hat{\varepsilon}, \alpha)}{k_0 \sqrt{\varepsilon_\perp}} B) = i \dfrac{\abs{\sin \alpha'}}{\tan \alpha'} \qty(A' + \dfrac{g(\hat{\varepsilon}', \alpha')}{k_0 \sqrt{\varepsilon'_\perp}} B'), \\
    E_z:\quad &i \abs{\sin \alpha} \qty(A - \dfrac{\abs{\cot \alpha}}{k_0 \sqrt{\varepsilon_\perp}} B) = - i \abs{\sin \alpha'} \qty(A' + \dfrac{\abs{\cot \alpha'}}{k_0 \sqrt{\varepsilon'_\perp}} B'), \\
    H_y:\quad &\dfrac{\sin \alpha \cos \alpha}{\abs{\cos \alpha}} \sqrt{\varepsilon_\perp} A + \dfrac{2 \abs{\sin \alpha} \varepsilon_\perp}{\varepsilon_{zy} k_0} B = - \dfrac{\sin \alpha' \cos \alpha'}{\abs{\cos \alpha'}} \sqrt{\varepsilon'_\perp} A' - \dfrac{2 \abs{\sin \alpha'} \varepsilon'_\perp}{\varepsilon'_{zy} k_0} B',
    \end{align*}
\end{widetext}
where \begin{equation*}
    g(\hat{\varepsilon}, \alpha) = \abs{\cot \alpha} + \dfrac{4 \varepsilon_\perp}{(\varepsilon_\parallel - \varepsilon_\perp) \abs{\sin 2 \alpha}}
\end{equation*}
and \( g(\hat{\varepsilon}', \alpha') \) are the same expression but with apostrophe sign.

The system can be rewritten as \( M (A A' B B')^\text{T} = 0 \) where \( M \) is a matrix \( 4 \times 4 \).
A non-trivial solution of the system exists if \( \det(M) = 0 \).
Choosing the signs of \( \alpha \) and \( \alpha' \) it can be reduced to \eqref{eq:dispersion-equation-diff} and \eqref{eq:dispersion-equation-same} after some rather cumbersome calculations.

The following formulas is derived from the system and determine the relative contributions of partial waves in the bisingular surface polariton
\begin{align*}
    \dfrac{A'}{A} &= \dfrac{\varepsilon_\perp}{\varepsilon'_\perp} = \cos^2 \varphi_\star, \\
    \dfrac{B'}{B} &= sp\,\dfrac{\eta' \sqrt{\varepsilon'_\perp}}{\eta \sqrt{\varepsilon_\perp}}\cdot \dfrac{(p - 1) \eta\cos^2\alpha + 2 (\cos^{-1}\varphi_\star + p)}{(1 - s p) \eta' \cos^2 \alpha' + 2 (1 + p \cos\varphi_\star)},\\
    \dfrac{A}{B} &= \dfrac{1}{\sqrt{\varepsilon_\perp} \tan\alpha} \cdot \dfrac{1 - B'/B}{1 + p \cos\varphi_\star},\\
    \dfrac{A'}{B'} &= \dfrac{1}{\sqrt{\varepsilon'_\perp} \tan\alpha} \cdot \dfrac{B'/B - 1}{\cos^{-1}\varphi_\star + p},\\
\end{align*}
where \( s \equiv \operatorname{sgn}{\alpha'} \) is the sign of \( \alpha' \), and \( p \equiv \abs{\tan \alpha' / \tan \alpha} \).
Here is taken into account that \( \alpha > 0 \).
Notice that \( 0 \leqslant p < 1 \).
It do not applicable if \( \alpha' = 0 \) or \( \eta, \eta' = 0 \).
The first case corresponds to nonlocalized wave in the upper medium \( \lambda' = 0 \) or \( \varphi = \varphi_\star \).
The latter two cases correspond to isotropic media.

\bibliography{bibliography.bib}

\end{document}